\documentclass{IEEEtran}
\usepackage{framed,multirow}
\usepackage{cite}
\usepackage{amsmath,amssymb,amsfonts}
\usepackage{algorithmic}
\usepackage{graphicx}
\usepackage{textcomp}
\usepackage{multirow}
\usepackage{comment}
\usepackage{tabu}
\usepackage{makecell}
\usepackage[colorlinks=true]{hyperref}

\def\etal{\emph{et al.}}
\def\ie{\emph{i.e.}}
\def\eg{\emph{e.g.}}

\begin{document}
\title{Image Synthesis with Disentangled Attributes for Chest X-Ray Nodule Augmentation and Detection}
\author{Zhenrong Shen, Xi Ouyang, Bin Xiao, Jie-Zhi Cheng, Qian Wang*, Dinggang Shen*, \IEEEmembership{Fellow, IEEE}
\thanks{Zhenrong Shen, Xi Ouyang and Bin Xiao are with the School of Biomedical Engineering, Shanghai Jiao Tong University, Shanghai, China. (e-mail: \{zhenrongshen, xi.ouyang, bin.xiao\}@sjtu.edu.cn).}
\thanks{Jie-Zhi Zheng is with the Department of Research and Development, Shanghai United Imaging Intelligence Co., Ltd., Shanghai, China. (e-mail: jiezhicheng@gmail.com).}
\thanks{Dinggang Shen and Qian Wang are with the School of Biomedical Engineering, ShanghaiTech University, Shanghai, China. (e-mail:\{dgshen, wangqian2\}@shanghaitech.edu.cn). Dinggang Shen is also with the Department of Research and Development, Shanghai United Imaging Intelligence Co., Ltd., Shanghai, China.}
}

\maketitle

\begin{abstract}
Lung nodule detection in chest X-ray (CXR) images is common to early screening of lung cancers. 
Deep-learning-based Computer-Assisted Diagnosis (CAD) systems can support radiologists for nodule screening in CXR. 
However, it requires large-scale and diverse medical data with high-quality annotations to train such robust and accurate CADs. 
To alleviate the limited availability of such datasets, lung nodule synthesis methods are proposed for the sake of data augmentation. 
Nevertheless, previous methods lack the ability to generate nodules that are realistic with the shape/size attributes desired by the detector. 
To address this issue, we introduce a novel lung nodule synthesis framework in this paper, which decomposes nodule attributes into three main aspects including shape, size, and texture, respectively.
A GAN-based Shape Generator firstly models nodule shapes by generating diverse shape masks. 
The following Size Modulation then enables quantitative control on the diameters of the generated nodule shapes in pixel-level granularity. 
A coarse-to-fine gated convolutional Texture Generator finally synthesizes visually plausible nodule textures conditioned on the modulated shape masks. 
Moreover, we propose to synthesize nodule CXR images by controlling the disentangled nodule attributes for data augmentation, in order to better compensate for the nodules that are easily missed in the detection task. 
Our experiments demonstrate the enhanced image quality, diversity, and controllability of the proposed lung nodule synthesis framework. 
We also validate the effectiveness of our data augmentation on greatly improving nodule detection performance.
\end{abstract}


\section{Introduction}
\IEEEPARstart{W}{hile} lung cancer is one of the leading healthcare threats worldwide with the highest morbidity and mortality among all the malignant tumors, early diagnosis and treatment can significantly increase the chance of surviving of the patients \cite{bray2018global,siegel2019cancer}. 
Lung nodules, which are typical manifestations of lung cancers, provide basis to radiological diagnoses even in early stages \cite{hansell2008fleischner}. 
Among all lung nodule screening tools, the chest X-ray (CXR) imaging contributes significantly due to its low dose and cost \cite{bhargavan2008trends}. 
However, the task of finding nodules from CXR images is difficult, even for expert radiologists \cite{shah2003missed}. An early retrospective study estimates that the lung nodules can be missed in 90$\%$ cases \cite{litjens2010simulation}. 

\begin{figure}[!t]
    \centering
    \includegraphics[scale=.5]{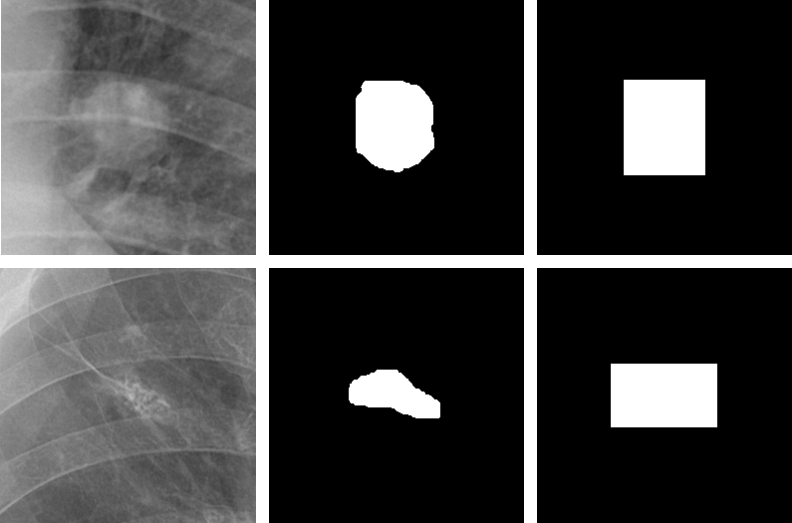}
    \caption{Two examples of lung nodule CXR patches and their corresponding masks. The columns from left to right refer to original patches, shape masks, and box masks, respectively. Previous studies only apply box masks as conditions while our method synthesizes lung nodules conditioned on shape masks.}
    \label{fig:mask_demo}
\end{figure}

To enhance the sensitivity of lung nodule screening using CXR images, several Computer-Assisted Diagnosis (CAD) systems based on deep learning have been developed in the past years \cite{nam2019development, li2020multi, sim2020deep}. 
In general, the CAD system usually requires a tremendous amount of high-quality annotated data that are diverse enough to cover a sufficient population distribution \cite{wang2021realistic}. 
But constructing such datasets poses a great challenge. 
First, the efforts in collecting extensive clinical data can hardly be neglected due to privacy concerns and reluctance to share data across medical institutes \cite{cunniff2000informed, yi2019generative, li2021medical}. 
Second, the annotation work is expensive and laborious as it requires high expertise due to the intrinsic difficulty of reading CXR images for lung nodules. 
These issues restrict the availability of sufficient data for supervised training. 
Moreover, traditional data augmentation techniques (\eg, image flipping, shifting, and rotation) bring limited improvement because the diversity of the training dataset is scarcely expanded \cite{shorten2019survey}. 
Thus, it is highly desired to synthesize lung nodule CXR images, which can effectively augment the data and then train the lung nodule detection models.

\begin{figure*}[ht]
    \centering
    \includegraphics[width=\textwidth]{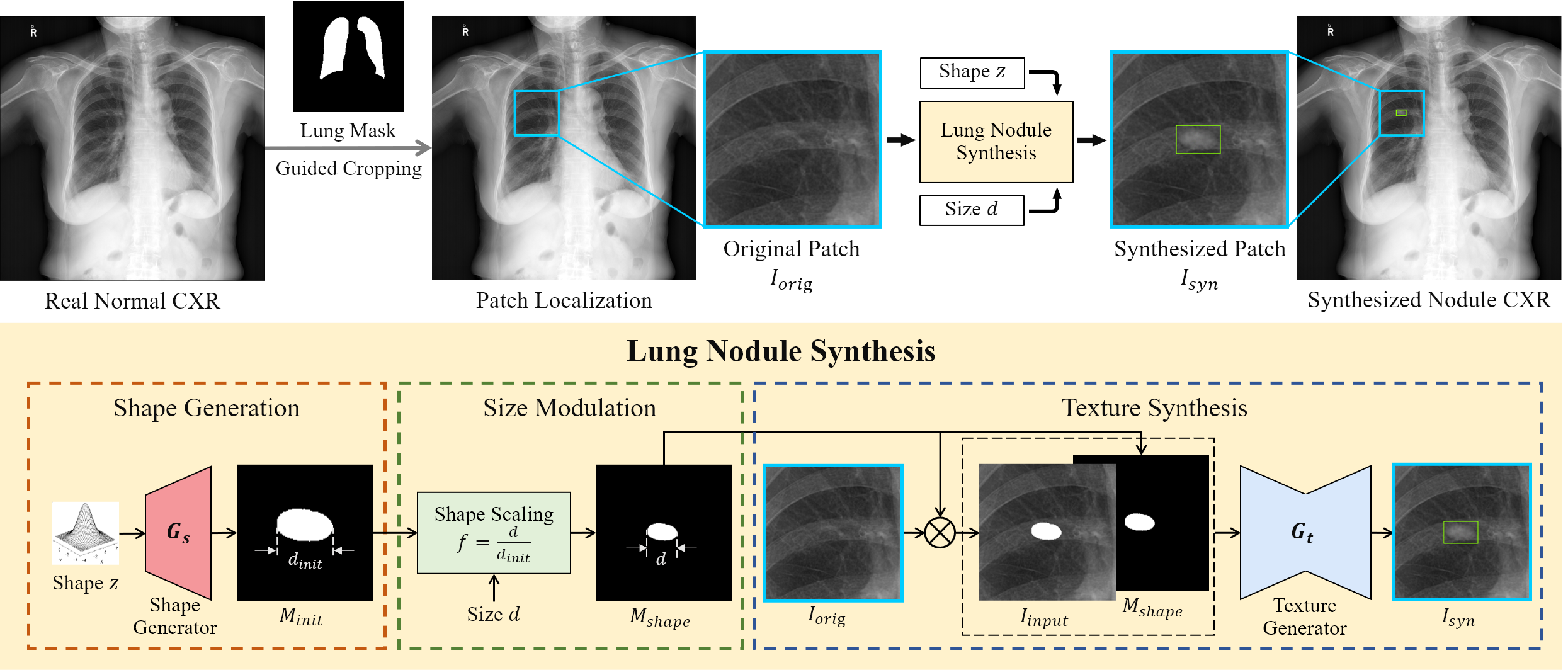}
    \caption{The overview of our lung nodule synthesis method. The lung mask guides the cropping of an original patch $I_{orig}$ from a real normal CXR image. $I_{orig}$, as well as a shape parameter $z$ and a size parameter $d$, is input to the proposed lung nodule synthesis framework to produce the synthesized patch $I_{syn}$ for obtaining a synthesized nodule CXR image. The lung nodule synthesis framework consists of three steps named Shape Generation, Size Modulation and Texture Synthesis, respectively.}
    \label{fig:pipeline}
\end{figure*}

It is challenging to synthesize visually plausible lung nodules in CXR images. 
First, the nodules targeted in clinical screening are often small, occupying only a few pixels to the minimum \cite{ann2021multi}. 
Second, the contexts surrounding the nodules are complex in CXR, especially after all tissues are projected and superimposed on a 2D image. 
To overcome these problems, recent studies \cite{gundel2020extracting, shen2021nodule, ann2021multi} propose several methods based on Generative Adversarial Networks (GANs) \cite{goodfellow2014generative} or image inpainting \cite{elharrouss2020image} to achieve realistic lung nodule synthesis. 
And the latest NODE21\footnote{\url{https://node21.grand-challenge.org/}} Challenge compares individual generative models in terms of their capability of improving nodule detection.
However, these works adopt box masks as indicated in Fig. \ref{fig:mask_demo}, which lack the fine granularity in perceiving and generating the nodules.

In this paper, in order to tackle the limitations above, we propose a novel lung nodule synthesis framework to allow quantitative control over the attributes of the synthesized nodules, \ie, shapes and sizes.
Formulated in an image inpainting scheme illustrated by Fig. \ref{fig:pipeline}, the proposed nodule synthesis framework disentangles the nodule attributes to model the shape, the size, and the texture, respectively. 
\begin{itemize}
    \item First, a GAN-based Shape Generator produces diverse shape masks (instead of rough box masks) that contour the nodule shapes. 
    \item Then, the diameters of the nodule shapes are modulated through an easy-to-use Size Modulation.
    \item Finally, a coarse-to-fine gated convolutional Texture Generator applies the modulated shape masks as the conditions to synthesize visually plausible nodule textures.
\end{itemize}

In addition, we design a Hard Example Mining (HEM) strategy \cite{shrivastava2016training} for data augmentation in subsequent nodule detection, as we can synthesize the nodules of the attributes that are specifically desired to improve the pre-trained detector.
The collaboration from the nodule synthesis framework improves the nodule detection performance, \eg, by augmenting the nodules that are easily missed. Extensive experiments are conducted to demonstrate the effectiveness of the proposed data synthesis and augmentation solution.

The rest of this paper is organized as follows. Section \ref{sec:related_works} reviews the related works. Section \ref{sec:method} presents the proposed lung nodule synthesis framework and our data augmentation strategy in detail. Section \ref{sec:results} explains the details of the dataset, the experimental setup, and the evaluation results for both lung nodule synthesis quality and data augmentation effectiveness. Finally, we conclude our work and discuss more in Section \ref{sec:conclusion}.

\section{Related Works}
\label{sec:related_works}
Since our lung nodule synthesis framework is mainly built in an image inpainting scheme, we herein provide detailed reviews of the related works in image inpainting as well as lung nodule synthesis in CXR images.

\subsection{Image Inpainting}
Image inpainting \cite{bertalmio2000image} refers to synthesizing visually seamless and semantically plausible image contents in the missing regions that are usually indicated by binary masks. Traditional methods \cite{efros2001image, kwatra2005texture, barnes2009patchmatch} rely on the principle of borrowing the most similar patches from known regions but usually fail to hallucinate realistic image contents when involving complicated scenes and non-repetitive structures. 

In recent years, a series of deep-learning-based methods \cite{elharrouss2020image} have been intensively developed to advance image inpainting. These approaches can synthesize reasonable contents by exploiting large-scale datasets and the semantics. 
Mainstream methods broadly fall into two categories: 
\begin{itemize}
    \item Enhancing vanilla convolutions. 
    Liu \etal \cite{liu2018image} introduced partial convolutions where the operations were conditioned on masked pixels only. 
    Gated convolution \cite{yu2019free} was then proposed to learn a dynamic feature selection mechanism. 
    Promising inpainting results for irregular corruptions are achieved by using these methods, but a common drawback of them relates to possible missing of fine structures of the inpainting outputs.
    
    \item Incorporating structural semantics. 
    The missing structures can be explicitly acquired in the beginning stage and further guide the texture generation in the following stage. 
    For example, EdgeConnect \cite{nazeri2019edgeconnect} and StructureFlow \cite{ren2019structureflow} exploited edge maps and edge-preserved smooth images, respectively.
    Though the fine details can be recovered, hallucinating reasonable structures itself is a challenging task and may suffer from large errors. 
    To further improve the inpainting performance, recent works attempt to fuse structural and textural features.
    For instance, PRVS \cite{li2019progressive} and MEDFE \cite{liu2020rethinking} mixed the modeling of structures and textures via a shared generator. 
    CTSDG \cite{guo2021image} facilitated more sufficient complements between structures and textures using a two-stream architecture. 
    These approaches employ sophisticated learning architectures to enhance the consistency between structures and textures, yet they may not be cost-effective in practice. 
\end{itemize}

Lung nodules usually appear as well-circumscribed or diffuse opacities in chest radiographs \cite{ost2012decision}.
Explicit and characteristic image structures hardly exist inside the nodule contours. 
In view of this fact, image inpainting methods that involve structural information in modeling are not suitable for generating realistic nodules.
In this paper, we utilize the modified version of convolutions to automatically grasp effective features from the surrounding contexts, which benefits plausible nodule texture synthesis most. 

\subsection{Lung Nodule Synthesis}
The development of CAD systems requires extensive and diverse CXR images as well as corresponding high-quality annotation. 
Concerning the very high cost to collect such data, the need for lung nodule synthesis in CXR images is naturally raised. Litjens \etal \cite{litjens2010simulation} simulated lung nodules in chest radiographs by using CT volumes containing real nodules. 
They projected CT data to generate nodule patches and then superimpose them to normal CXR images. 

Deep-learning-based methods have been developed recently to gain more flexibility in generating lung nodules. 
The first attempt to generate CXR images with multiple abnormalities was investigated by Salehinejad \etal \cite{salehinejad2018synthesizing}. DCGANs \cite{radford2015unsupervised} were applied for each thoracic disease independently to synthesize whole CXR images for tackling the class imbalance problem in the image-level classification task. 
However, the presence and the location of the target disease cannot be manually controlled. 
To overcome this issue, Ann \etal \cite{ann2021multi} proposed a PGGAN-based \cite{karras2017progressive} model conditioned on box masks.
Nevertheless, this approach has only achieved low-resolution ($256\times256$) results that hardly preserve local details for small-size lung nodules. 
Therefore, generating a whole image directly is not suitable for synthesizing high-resolution CXR images with subtle lung nodules. 

Another technical route for deep-learning-based lung nodule synthesis is image inpainting. 
Image inpainting can edit a local region only given its surroundings in a patch and keep the original spatial resolution unchanged, which fits high-resolution CXR images perfectly. 
Sogancioglu \etal \cite{sogancioglu2018chest} firstly investigated the performance of several image inpainting models \cite{pathak2016context, yeh2017semantic, yu2018generative} applied to CXR images, and demonstrated the feasibility of generating realistic and seamless patches based on image inpainting. 
Gundel \etal \cite{gundel2020extracting} then proposed a local feature augmentation method. They extracted isolated nodules from the residuals between real nodule patches and their inpainted normal-looking counterparts, and then blended them into real normal CXR images for data augmentation. 
However, this method only executes a copy-and-paste work, resulting in the limited variety of the synthesized data. 

In contrast to the studies above, our previous work \cite{shen2021nodule} employed a partial convolutional \cite{liu2018image} U-Net \cite{ronneberger2015u} to directly synthesize lung nodules and utilized a classifier to select the synthesized hard samples for effective data augmentation on lung nodule detection. 
In this study, to further enhance the image quality, the diversity, the controllability, and the data augmentation effectiveness, we build a coarse-to-fine gated convolutional network conditioned on a controllable nodule shape mask and design an HEM-based data augmentation strategy.

\section{Method}
\label{sec:method}
In this paper, we propose to disentangle the overall lung nodule synthesis into three steps of Shape Generation, Size Modulation, and Texture Synthesis. 
As demonstrated in Fig. \ref{fig:pipeline}, the overview of the proposed method is presented as follows.
\begin{enumerate}
    \item A pre-trained FCDenseNet \cite{jegou2017one} segments the lung mask from a real normal CXR image. 
    A local patch $I_{orig}$ within the lungs is then cropped from the CXR image under the guide of the lung mask. 
    
    \item The Shape Generator $G_{s}$ produces an initial shape mask $M_{init}$ via the shape parameter $z$ sampled from a standard Gaussian latent space. 
    
    \item The size of the initial shape mask in $M_{init}$ is scaled from the initial diameter $d_{init}$ to a desired diameter $d$ by the factor $f=\frac{d}{d_{init}}$, resulting in a modulated shape mask $M_{shape}$. 
    
    \item The original patch $I_{orig}$ and the modulated shape mask $M_{shape}$ compose the input to the Texture Generator $G_{t}$.
    The Texture Generator $G_{t}$ synthesizes visually realistic nodule texture inside the shape region of $M_{shape}$. 
    
    \item Finally, we put the synthesized nodule patch $I_{syn}$ in the place of $I_{orig}$ to obtain the synthesized nodule CXR image. 
\end{enumerate}

\begin{figure}[ht]
    \centering
    \includegraphics[scale=0.5]{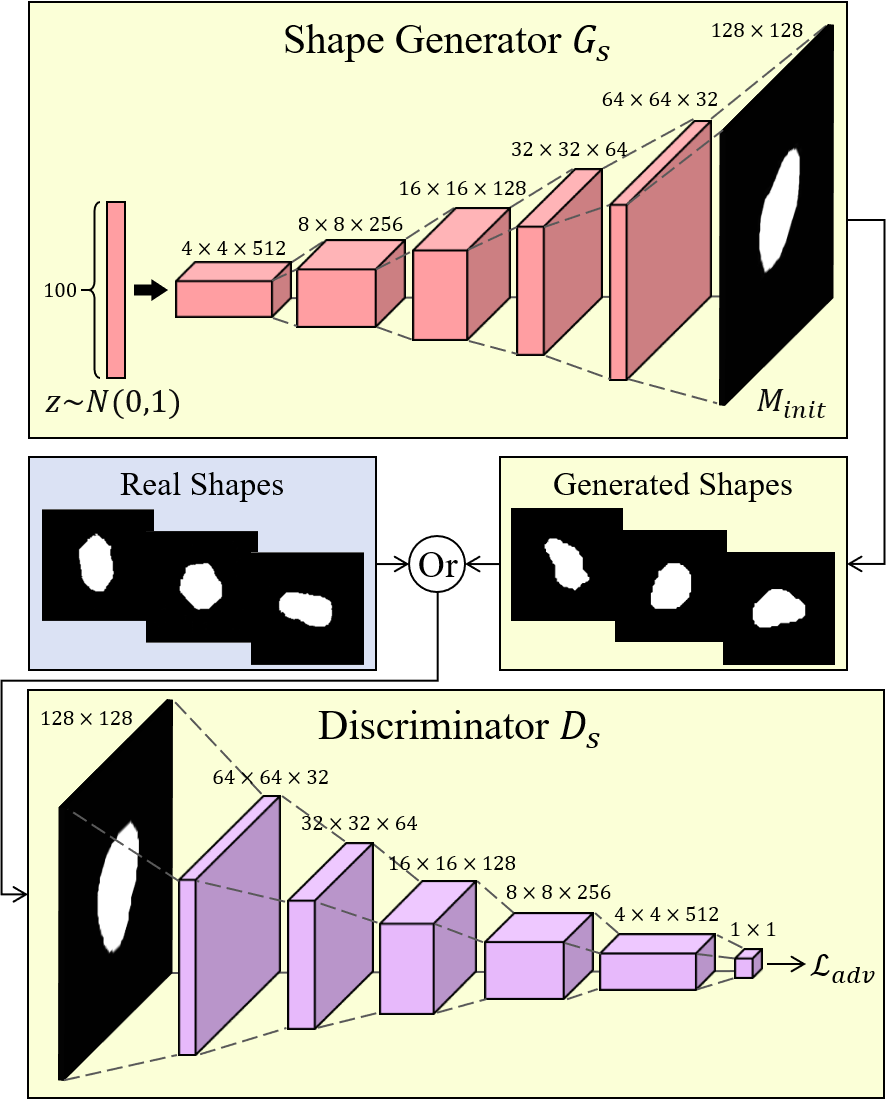}
    \caption{The detailed structure of the proposed Shape Generator in the \textit{Shape Generation} step. A DCGAN architecture, which is composed of a generator $G_{s}$ and a discriminator $D_{s}$, is adopted in training the Shape Generator to learn the data distribution of lung nodule shapes.}
    \label{fig:shape_generator}
\end{figure}

\subsection{Nodule Shape Generation}
To make sure the shape of the generated nodule matches the distribution of real nodules, we consider the nodule shapes explicitly in generating conditional shape masks. 
A shape mask is defined as a binary image where 1 and 0 stand for the shape foreground and background, respectively. 
As illustrated in Fig. \ref{fig:shape_generator}, we use a DCGAN architecture to train the proposed Shape Generator $G_s$.
The trained $G_s$ learns to map a Gaussian distribution $p(z)\sim N(0,I)$ to the underlying distribution of real nodule shapes. 
To be specific, a 100-dimensional latent vector $z$ is randomly sampled from $N(0,I)$ and is spatially extended to a $4\times4\times512$ feature map. 
Five $4\times4$ transposed convolutional layers with upsampling stride 2, which are followed by batch normalizations and ReLU activations, convert the projected feature map into a $128\times128$ binary mask $M_{init}$. 

The discriminator $D_s$, which echos $G_s$ in the architecture, tells whether an input mask is the generated shape mask $M_{init}$ or the real shape mask $M_{gt}$. 
By competing with each other in an adversarial setting, the Shape Generator $G_s$ is optimized to deceive the discriminator $D_s$ so that $D_s$ cannot easily distinguish the generated shapes from the real ones. 
To alleviate the mode collapse problem in regular GANs, we adopt the LSGAN loss \cite{mao2017least} in training $G_s$, which is represented as:

\begin{equation}
    \begin{aligned}
    \min_{D_s}{\mathcal{L}_{adv}(D_s)}=&\frac{1}{2}\mathbb{E}_{M_{gt}}\left[\left(D_s\left(M_{gt}\right)-1\right)^2\right] \\
    &+\frac{1}{2}\mathbb{E}_{z\sim p(z)}\left[\left(D_s\left(G_s\left(z\right)\right)\right)^2\right],
    \end{aligned}
\end{equation}
\begin{equation}
    \min_{G_s}{\mathcal{L}_{adv}(G_s)}=\frac{1}{2}\mathbb{E}_{z\sim p(z)}\left[\left(D_s\left(G_s\left(z\right)\right)-1\right)^2\right].
\end{equation}

In view of the fact that the real shape masks are unbalanced in size, we normalize the diameters of all ground-truth shape masks to the same $d_{init}$ when using them to train $G_s$. This is achieved by the Size Modulation, which will be introduced in the next subsection. 

\subsection{Nodule Size Modulation}
Advanced GAN-based image generation methods can control the variation of high-level  semantic attributes (\eg, age, gender, and expression) in the image space by manipulating the latent codes in the latent space \cite{goetschalckx2019ganalyze, shen2020interpreting}. 
Unlike natural images, the pixel/voxel spacing in medical images can reflect actual physical sizes of the objects in the real world. 
Therefore, the physical size of a lesion can be precisely controlled at the pixel/voxel-level granularity by simple morphological operation in the image space instead of complicated image encoding and manipulation in the latent space. 
Motivated by this idea, we develop the easy-to-use Size Modulation based on image scaling to quantitatively control the diameters of the nodule shapes.

Specifically, the proposed Size Modulation consists of following steps. 
First, the initial shape mask $M_{init}$ is upsampled from $128\times128$ to $256\times256$ using the nearest-neighbor interpolation in order to match the patch size of the input to the Texture Generator. 
Second, the initial diameter $d_{init}$ of the shape region is calculated by averaging the lengths of major and minor axes of the ellipse which has the same normalized second central moments as the shape region. 
Third, the shape region image, which has the same size with the bounding box that tightly encloses the shape region, is cropped from the upsampled $M_{init}$. 
The cropped image is then scaled by a certain factor computed as the ratio of a user-input diameter $d$ to the original one $d_{init}$. 
Finally, the conditional shape mask $M_{shape}$ is obtained by overlaying the rescaled shape region image on the center of a newly created $256\times256$ null matrix.

\begin{figure*}[ht]
    \centering
    \includegraphics[width=\textwidth]{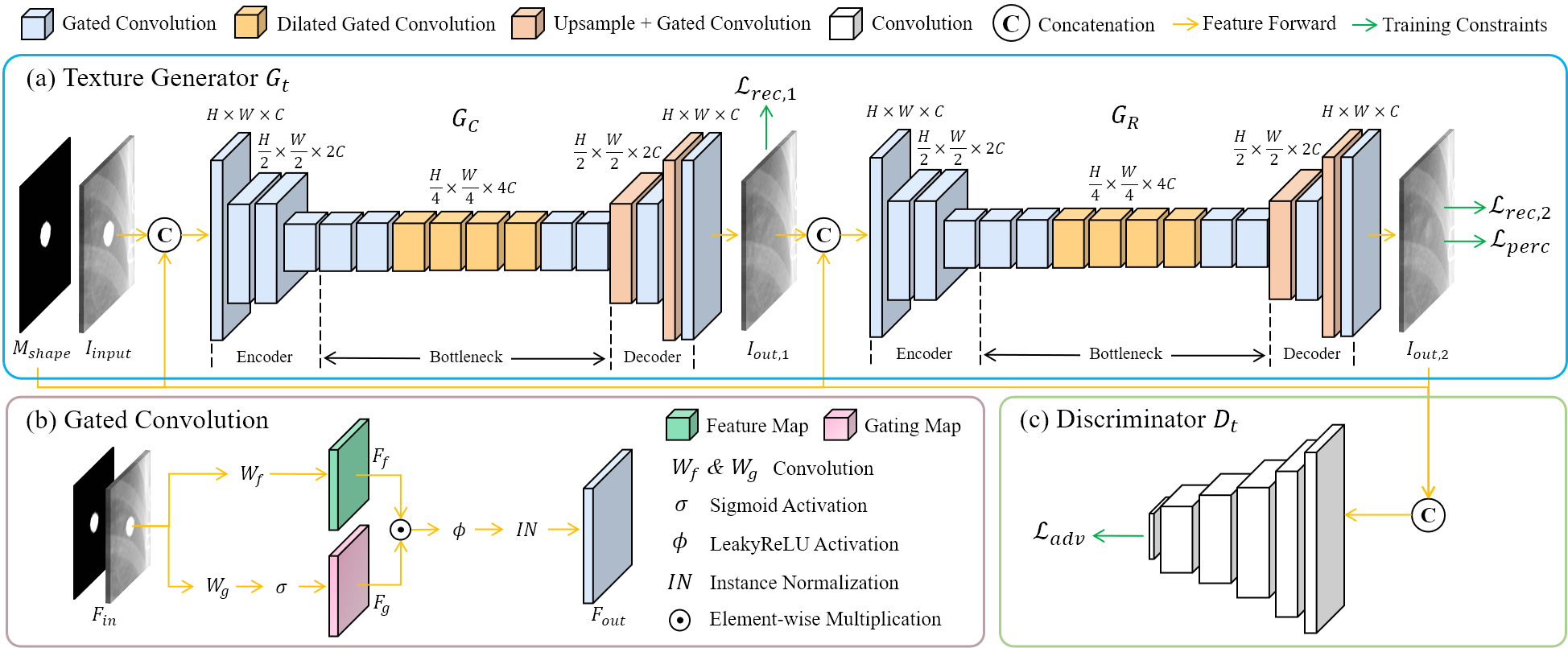}
    \caption{The architecture of the proposed Texture Generator in the \textit{Texture Synthesis} step. This generator is equipped with gated convolutions and consists of a coarse generator $G_C$ and a refinement generator $G_R$. The whole network is optimized with the reconstruction loss $\mathcal{L}_{rec}$, the perceptual loss $\mathcal{L}_{perc}$, and the adversarial loss $\mathcal{L}_{adv}$ given by the discriminator $D_t$.}
    \label{fig:texture_synthesis}
\end{figure*}

\subsection{Nodule Texture Synthesis}
The dynamic feature selection mechanism of gated convolution \cite{yu2019free} makes it suitable for lung nodule synthesis because it automatically learns the optimal feature extraction of contextual information from the given data. 
As shown in Fig. \ref{fig:texture_synthesis}(b), the gated convolution can be formulated as:
\begin{equation}
    \begin{aligned}
        F_{out}&=IN\left(\phi\left(F_f\odot F_g\right)\right) \\
        &=IN\left(\phi\left(\left(W_f\cdot F_{in}\right)\odot \left(\sigma\left(W_g\cdot F_{in}\right)\right)\right)\right),
    \end{aligned}
\end{equation}
where $\sigma$, $\phi$, and \textit{IN} denote sigmoid activation, LeakyReLU activation, and instance normalization, respectively. 
For an input feature map $F_{in}$, a vanilla convolution kernel $W_f$ extracts its intrinsic features $F_f$ while the other one $W_g$ is used to obtain the soft gating weights for each location. 
The gating map $F_g$ is then assigned across $F_f$ to gain effective features that contribute to nodule synthesis. 

As shown in Fig. \ref{fig:texture_synthesis}(a), we adopt gated convolutions to build the Texture Generator $G_t$ in a coarse-to-fine manner, which synthesizes nodule textures inside the shape regions marked by the conditional shape masks. 
The input patch is formulated as $I_{input}=I_{orig}\odot\left(\mathbf{1}-M_{shape}\right)+M_{shape}$, where $\mathbf{1}$ has all-one elements in the same shape with $M_{shape}$. 
The conditional shape mask $M_{shape}$ is then concatenated to the input patch $I_{input}$, resulting in a 4-channel input tensor. 
Both the coarse generator $G_C$ and the refinement generator $G_R$ employ the same encoder-bottleneck-decoder architectures equipped with gated convolutional layers.
LeakyReLU activation and instance normalization are used across all the gated convolutional layers. 
The encoder downsamples the input tensor twice followed by the bottleneck to extract deep features. 
The symmetrical decoder upsamples the feature map to the original patch size via the nearest-neighbor interpolation. 
In order to gain a larger receptive field, four dilated gated convolutional layers with dilation factors \{2, 4, 8, 16\} are adopted in the middle of the bottleneck. 
The first output $I_{out,1}$ yielded by $G_C$ is stacked with $M_{shape}$ again to formulate a new input to $G_R$ for obtaining the final output $I_{out,2}$ (corresponding to $I_{syn}$ in Fig. \ref{fig:pipeline}) with finer texture details. 

A six-layer fully convolutional PatchGAN \cite{isola2017image} discriminator $D_t$ is employed to distinguish the genuine nodule patches from the synthesized ones. 
As displayed in Fig. \ref{fig:texture_synthesis}(c), the input tensor consists of the conditional shape mask $M_{shape}$ and the final output $I_{out,2}$, while the output feature map is a 1-channel feature map downsampled to $\frac{1}{32}$ of the original patch size. 
Spectral normalization \cite{miyato2018spectral} is used across all the convolutional layers as it can alleviate the training instability problem.

The Texture Generator $G_t$ is trained jointly with the reconstruction loss, the perceptual loss \cite{johnson2016perceptual} and the adversarial loss, to progressively render visually realistic results. 
\begin{itemize}
    \item \textbf{Reconstruction loss}. A pixel-wise $\ell_1$ reconstruction loss is adopted for both  $I_{out,1}$ and  $I_{out,2}$ to reduce their pixel intensity differences with the ground-truth $I_{gt}$:
    \begin{equation}
        \mathcal{L}_{rec,1}=\mathbb{E}\left[{\left\|I_{gt}-I_{out,1}\right\|}_{\ell_1}\right],
    \end{equation}
    \begin{equation}
        \mathcal{L}_{rec,2}=\mathbb{E}\left[{\left\|I_{gt}-I_{out,2}\right\|}_{\ell_1}\right].
    \end{equation}
    
    \item \textbf{Perceptual loss}. The perceptual loss is introduced to shorten the high-level contextual discrepancy between $I_{out,2}$ and $I_{gt}$ in the feature space. 
    It computes the $\ell_1$ distance of their feature representations by projecting them via a VGG-16 \cite{simonyan2014very} network pre-trained on ImageNet \cite{deng2009imagenet}:
    \begin{equation}
        \mathcal{L}_{perc}=\mathbb{E}\left[\sum_{i}{{\left\|\psi_i\left(I_{gt}\right)-\psi_i\left(I_{out,2}\right)\right\|}_{\ell_1}}\right],
    \end{equation}
    where $\psi_i$ denotes the activation map of the $i$-th pooling layer from VGG-16, and \textit{pool1}, \textit{pool2} and \textit{pool3} are used to extract the features of $I_{gt}$ and $I_{out,2}$ in our experiments.
    
    \item \textbf{Adversarial loss}. The adversarial loss enforces $G_t$ to generate visually plausible nodule textures that can deceive discriminator $D_t$ successfully. 
    Here we apply the LSGAN loss \cite{mao2017least} as the adversarial loss:
    \begin{equation}
        \begin{aligned}
            \min_{D_t}{\mathcal{L}_{adv}(D_t)}&=\frac{1}{2}\mathbb{E}_{I_{gt},M_{gt}}\left[\left(D_t\left(I_{gt},M_{gt}\right)-1\right)^2\right] \\
            &+\frac{1}{2}\mathbb{E}_{I_{out,2},M_{gt}}\left[\left(D_t\left(I_{out,2},M_{gt}\right)\right)^2\right],
        \end{aligned}
    \end{equation}
    \begin{equation}
        \begin{aligned}
            &\min_{G_t}{\mathcal{L}_{adv}(G_t)} \\
            &=\frac{1}{2}\mathbb{E}_{I_{out,2},M_{gt}}\left[\left(D_t\left(I_{out,2},M_{gt}\right)-1\right)^2\right] \\
            &=\frac{1}{2}\mathbb{E}_{I_{input},M_{gt}}\left[\left(D_t\left(G_t\left(I_{input},M_{gt}\right),M_{gt}\right)-1\right)^2\right].
        \end{aligned}
    \end{equation}
\end{itemize}

The total objective is formulated by combining all the loss functions above:
\begin{equation}
    \mathcal{L}_{total}=\lambda_{rec,1}\mathcal{L}_{rec,1}+\lambda_{rec,2}\mathcal{L}_{rec,2}+\lambda_{perc}\mathcal{L}_{perc}+\lambda_{adv}\mathcal{L}_{adv},
\end{equation}
where $\lambda_{rec,1}$, $\lambda_{rec,2}$, $\lambda_{perc}$ and $\lambda_{adv}$ represent weighting factors for different loss terms. 
All these factors are empirically set to 1 in our experiments.

\begin{figure*}[ht]
    \centering
    \includegraphics[width=0.9\textwidth]{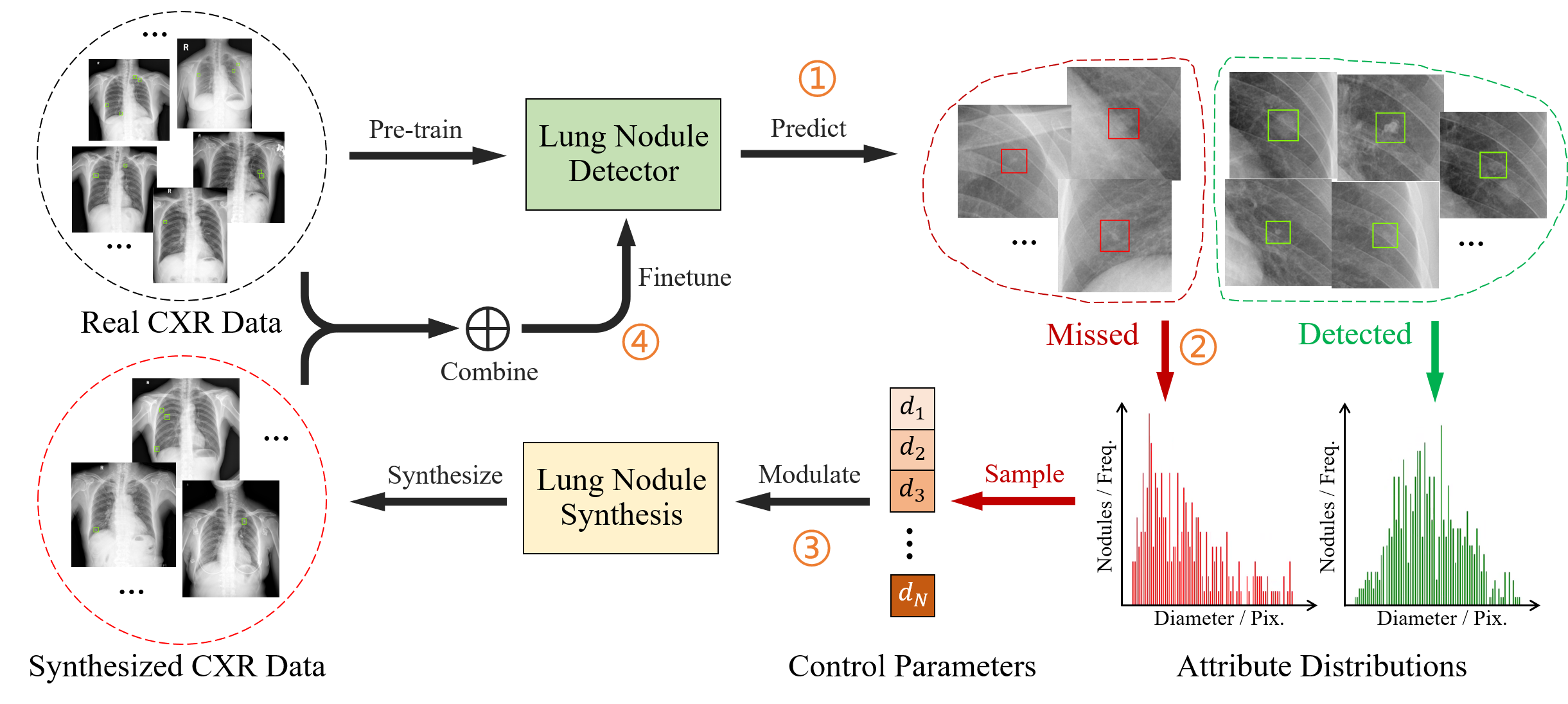}
    \caption{The proposed HEM-based data augmentation strategy consists of four steps: (1) A pre-trained lung nodule detector predicts detection results that are categorized as detected nodules and missed ones, respectively. (2) The attribute distributions (\eg, size distributions) are established from the detection results. (3) $N$ control parameters (\eg, diameters for size distribution) are sampled from the attribute distribution of missed nodules and then modulate lung nodule synthesis. (4) The synthesized CXR data are combined with real ones to finetune the detector.}
    \label{fig:data_aug}
\end{figure*}

\subsection{HEM-based data augmentation}
Our preliminary study \cite{shen2021nodule} illustrates that randomly adding synthesized data limits the data augmentation effectiveness while selecting hard examples deliberately can better improve the detection performance.
Thus, we propose the HEM-based data augmentation strategy involving the controllability of our lung nodule synthesis framework in order to achieve an enhanced data augmentation effectiveness.

As depicted in Fig. \ref{fig:data_aug}, the proposed data augmentation strategy mainly consists of four steps. 
\begin{enumerate}
    \item A pre-trained lung nodule detector firstly provides the detection results on a certain number of real CXR images.
    \item The results are grouped into the detected nodules and the missed ones.
    An explicit distribution of a particular nodule attribute is established from those missed nodules.
    \item We synthesize nodule images given the attribute distribution for the desired attributes. 
    \item The synthesized data are combined with the real ones to finetune the pre-trained detector for improving detection performance.  
\end{enumerate}

In this paper, we use the nodule size to demonstrate the attribute controllability of our nodule synthesis framework, as well as its contribution to effective data augmentation for nodule detection.
As shown in Fig. \ref{fig:data_aug}, we calculate the histogram of the missed nodules in terms of their sizes.
Several size parameters $d_1$, $d_2$, $\cdots$, $d_N$ (corresponding to diameters) are uniformly sampled from an array containing the diameters of each missed nodule, thus following the established size distribution.
These sampled control parameters are then used to modulate the sizes of the synthesized nodules via the Size Modulation.

\section{Experimental Results}
\label{sec:results}
\subsection{Dataset and Experimental Setup}
In this work, we involve 12,737 frontal-view CXR images collected from our collaborative hospitals, including 3,024 CXR images that are diagnosed as nodules and the other 9,713 images as normal cases. 
All the nodules are well-annotated with bounding boxes and shape contours by two experienced radiologists. 
We randomly select 2,420 nodule CXR images as the training set and the other 604 images as the testing set.
For each nodule CXR image, $256\times256$ nodule patches centered on the box annotations are extracted and their corresponding shape masks are derived from the shape annotations. 
All the experiments are conducted using an NVIDIA GeForce RTX 3090 GPU with the PyTorch framework \cite{paszke2019pytorch}.
The implementation details for each part of the experiments are presented as follows.

\begin{figure*}[ht]
    \centering
    \includegraphics[width=\textwidth]{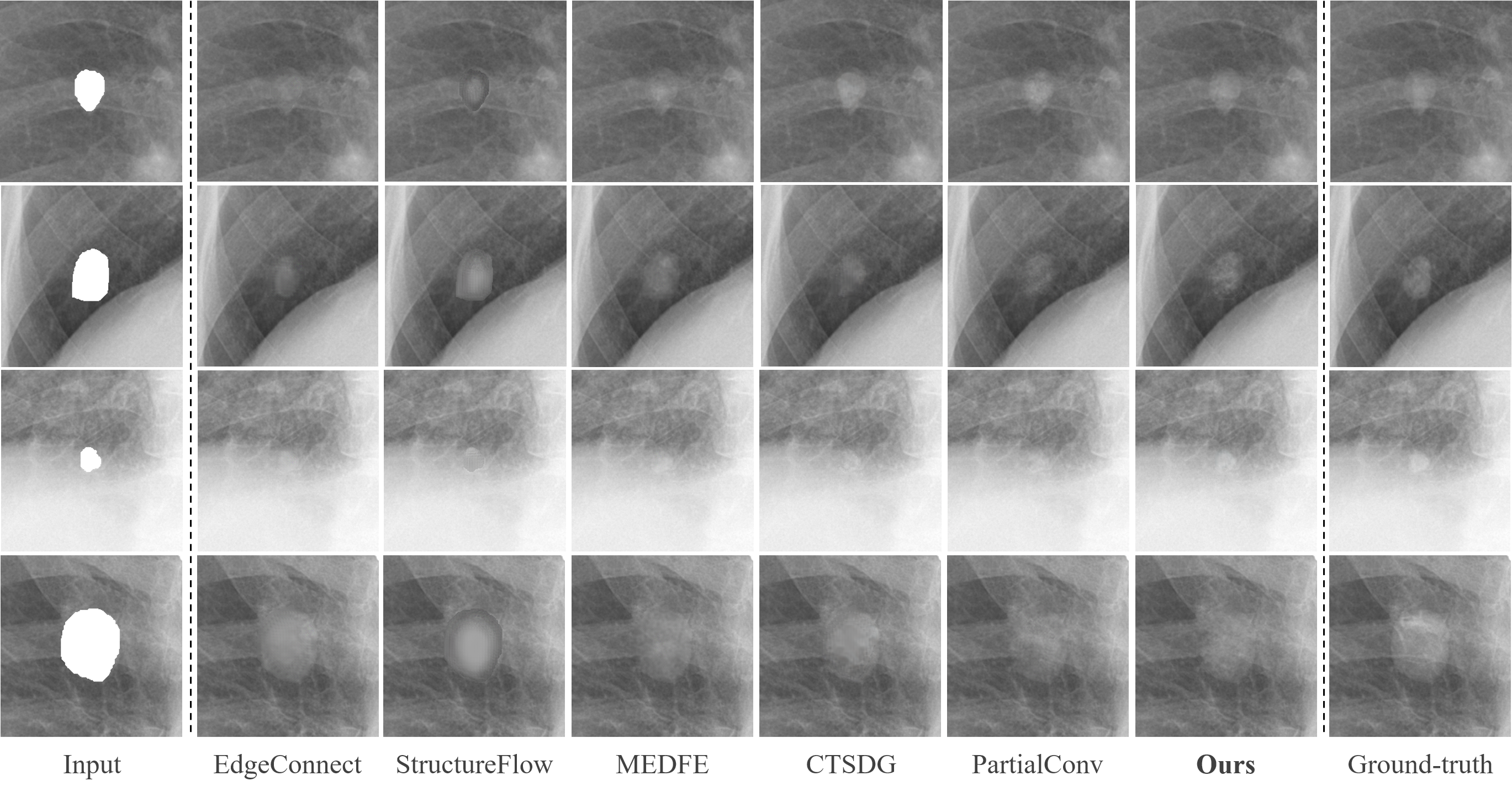}
    \caption{Qualitative comparison between state-of-the-art methods and the proposed method for 4 nodule patch cases (presented in rows). Columns from left to right stand for the inputs, the outputs by different methods, and the ground-truth.}
    \label{fig:syn_compare}
\end{figure*}

\begin{table*}[htbp]
\centering
\caption{Quantitative comparison between state-of-the-art methods and the proposed method (↓: Lower is better; ↑: Higher is better).}
\label{tab:syn_compare}
\setlength{\tabcolsep}{3mm}{
\begin{tabular}{c|llll|lll}
\hline
\multirow{2}{*}{Method} & \multicolumn{4}{c|}{Full Patch} & \multicolumn{3}{c}{Masked Region} \\ \cline{2-8} 
 & \multicolumn{1}{c}{MAE↓} & \multicolumn{1}{c}{PSNR↑} & \multicolumn{1}{c}{SSIM↑} & \multicolumn{1}{c|}{FID↓} & \multicolumn{1}{c}{MAE↓} & \multicolumn{1}{c}{PSNR↑} & \multicolumn{1}{c}{SSIM↑} \\ \hline \hline
EdgeConnect \cite{nazeri2019edgeconnect} & 0.0007 & 46.4647 & 0.9898 & 2.2673 & 0.0318 & 28.4179 & 0.7314 \\
StructureFlow \cite{ren2019structureflow} & 0.0016 & 39.4209 & 0.9624 & 23.2486 & 0.0826 & \multicolumn{1}{c}{21.3093} & \multicolumn{1}{c}{0.4616} \\
MEDFE \cite{liu2020rethinking} & 0.0008 & 46.3337 & 0.9892 & 2.3105 & 0.0324 & 28.2221 & 0.7204 \\
CTSDG \cite{guo2021image} & 0.0008 & 46.8018 & 0.9894 & 2.3398 & 0.0305 & 28.6902 & 0.7418 \\
PartialConv \cite{liu2018image} & 0.0010 & 45.8891 & 0.9884 & 2.5729 & 0.0325 & 28.1230 & 0.7187 \\ \hline
\textbf{Ours} & \multicolumn{1}{c}{\textbf{0.0007}} & \multicolumn{1}{c}{\textbf{47.0932}} & \multicolumn{1}{c}{\textbf{0.9903}} & \textbf{2.1268} & \textbf{0.0296} & \textbf{28.9815} & \textbf{0.7555} \\ \hline
\end{tabular}}
\end{table*}

\begin{itemize}
    \item \textbf{Shape Generator}. 
    The initial diameters $d_{init}$ of all ground-truth shape masks are normalized to 100-pixel in length using the Size Modulation firstly, and all these shape masks are then resized to $128\times128$ for training. 
    Both the Shape Generator $G_s$ and the discriminator $D_s$ are optimized using Adam optimizer \cite{kingma2014adam} with ${\beta}_1 =0.5$ and ${\beta}_2 =0.999$. 
    $G_s$ is trained with a learning rate of $10^{-4}$ for 1,000 epochs while the learning rate for $D_s$ is set to $10^{-5}$. 
    The batch size is set to 6.
    
    \item \textbf{Texture Generator}. 
    Adam optimization \cite{kingma2014adam} with ${\beta}_1 =0.5$ and ${\beta}_2 =0.999$ is adopted in training the Texture Generator $G_t$ and the discriminator $D_t$ with a batch size of 8. 
    We firstly train $G_t$ with an initial learning rate of $10^{-4}$ until convergence. 
    We then lower the learning rate to $10^{-5}$ and continue to train $G_t$ until another convergence. 
    And $D_t$ is trained with a learning rate set to one-tenth of $G_t$'s all the time. 
    
    \item \textbf{Data Augmentation}. 
    To prove the data augmentation effectiveness for typical one-stage and two-stage detection models, a Faster R-CNN \cite{ren2015faster} and a RetinaNet \cite{lin2017focal} with ResNet-50 \cite{he2016deep} backbones are applied for lung nodule detection. 
    The batch sizes are set to 2 while the rest hyperparameters adopt the original settings by Ren \etal \cite{ren2015faster} and Lin \etal \cite{lin2017focal}. 
    We conduct five-fold cross validation to achieve a comprehensive evaluation.
    For each fold, the detectors are firstly pre-trained using real nodule CXR images (the training set) with the learning rates of 0.001 for 20 epochs.
    Different combinations of real and synthesized data are then utilized to finetune the detectors with the learning rates of 0.0001 for 10 epochs.
    The augmented detectors are finally evaluated using real nodule CXR images (the testing set).
    The Intersection over Union (IoU) thresholds are set to 0.2 for all the detectors during the testing phase.
    
\end{itemize} 

\subsection{Nodule Synthesis Evaluation}
\subsubsection{Comparison with SOTA Methods}
Fig. \ref{fig:syn_compare} qualitatively compares several synthesized results by the Texture Generator $G_t$ with the state-of-the-art (SOTA) image inpainting methods. 
EdgeConnect \cite{nazeri2019edgeconnect} and StructureFlow \cite{ren2019structureflow} fail in generating seamless patterns inside the masked regions. 
The results of MEDFE \cite{liu2020rethinking} tend to be blurred near the ribs. Obvious artifacts can be seen in the outputs of CTSDG \cite{guo2021image} if the masked regions are large. 
PartialConv \cite{liu2018image}, which is implemented in our previous work \cite{shen2021nodule}, is suitable for synthesizing nodules, but is still inadequate in generating finer details when meeting complex scenes. 
For example, compared with the output of our method (Row 7, Column 3), the result of PartialConv (Row 6, Column 3) has lower contrast when the nodule is near the diaphragm and the spine, which is easily confused with the background. 
In summary, our method is able to synthesize more visually realistic and contextually robust lung nodules regardless of the variation of the surrounding anatomical structures and the sizes of masked regions.

To quantitatively compare the performance between our method and the SOTA image inpainting methods, we use Peak Signal to Mean Absolute Error (MAE), Noise Ratio (PSNR), Structural Similarity (SSIM), and Fr\'echet Inception Distance (FID) \cite{heusel2017gans} as evaluation metrics for the full patches and the masked shape regions. 
As shown in Table \ref{tab:syn_compare}, StructureFlow achieves the worst scores, which is consistent with its visualization results. 
Though PartialConv generates visually plausible results, its scores are worse than most other methods. 
Our method outperforms all the other approaches in both qualitative and quantitative comparisons, demonstrating its superiority in synthesizing realistic lung nodules.

\begin{figure*}[ht]
    \centering
    \includegraphics[width=\textwidth]{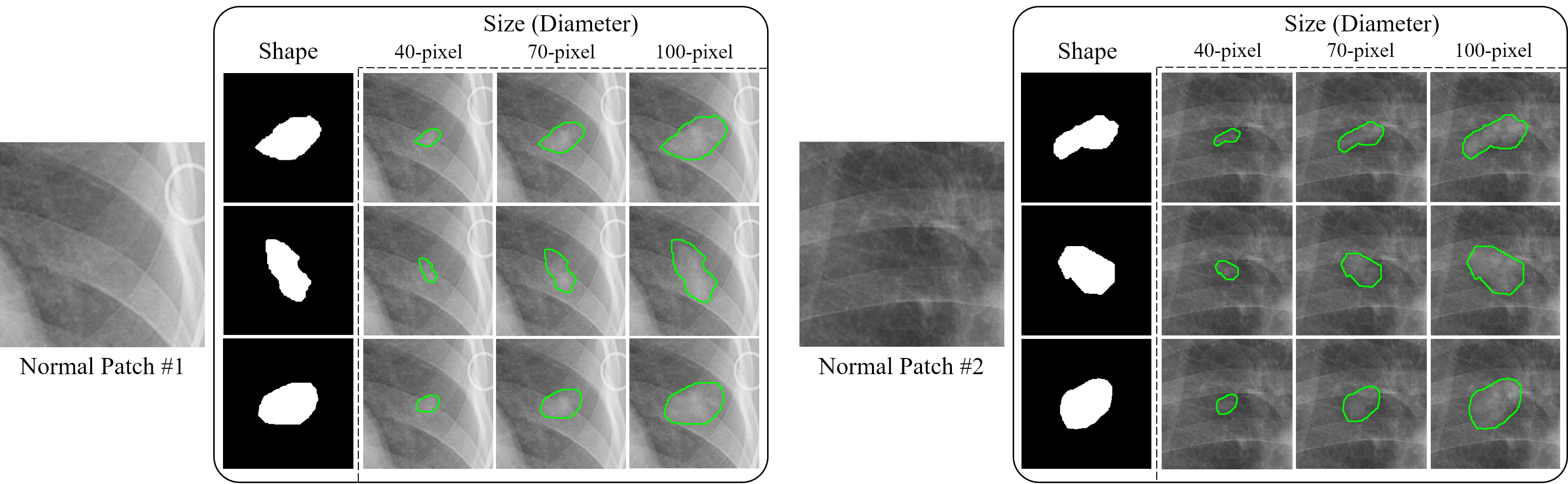}
    \caption{Two examples to show the controllability of nodule shapes and sizes for the synthesized nodules. Random combinations between the shapes and the sizes can synthesize lung nodules that do not exist in the real world but match the distribution of real nodules.}
    \label{fig:control}
\end{figure*}

\subsubsection{Controllability of Shapes and Sizes}
The diversity of the shapes and the sizes of the synthesized lung nodules can be controlled by the combinations of different conditional shape masks and various user-input diameters. 
The experimental results are shown in Fig. \ref{fig:control}.
The normal patches are randomly cropped from our normal CXR image dataset. 
All the conditional shape masks are generated using the proposed Shape Generator. 
The visual plausibility of these shape masks demonstrates that our Shape Generator faithfully learns what a real nodule should look like in shape.
For each shape mask, the Size Modulation algorithm scales the diameters of nodule shapes to 40-pixel, 70-pixel, and 100-pixel in length, respectively.
As illustrated by the green contours, the sizes of nodules can be modulated without changing nodule shapes.
By combining the normal patches with different modulated shape masks, the Texture Generator automatically replenishes the shape regions with realistic nodule textures, thus providing diverse lung nodule samples that do not exist in the real world but match the distribution of real nodules.

\begin{figure*}[ht]
    \centering
    \includegraphics[width=0.75\textwidth]{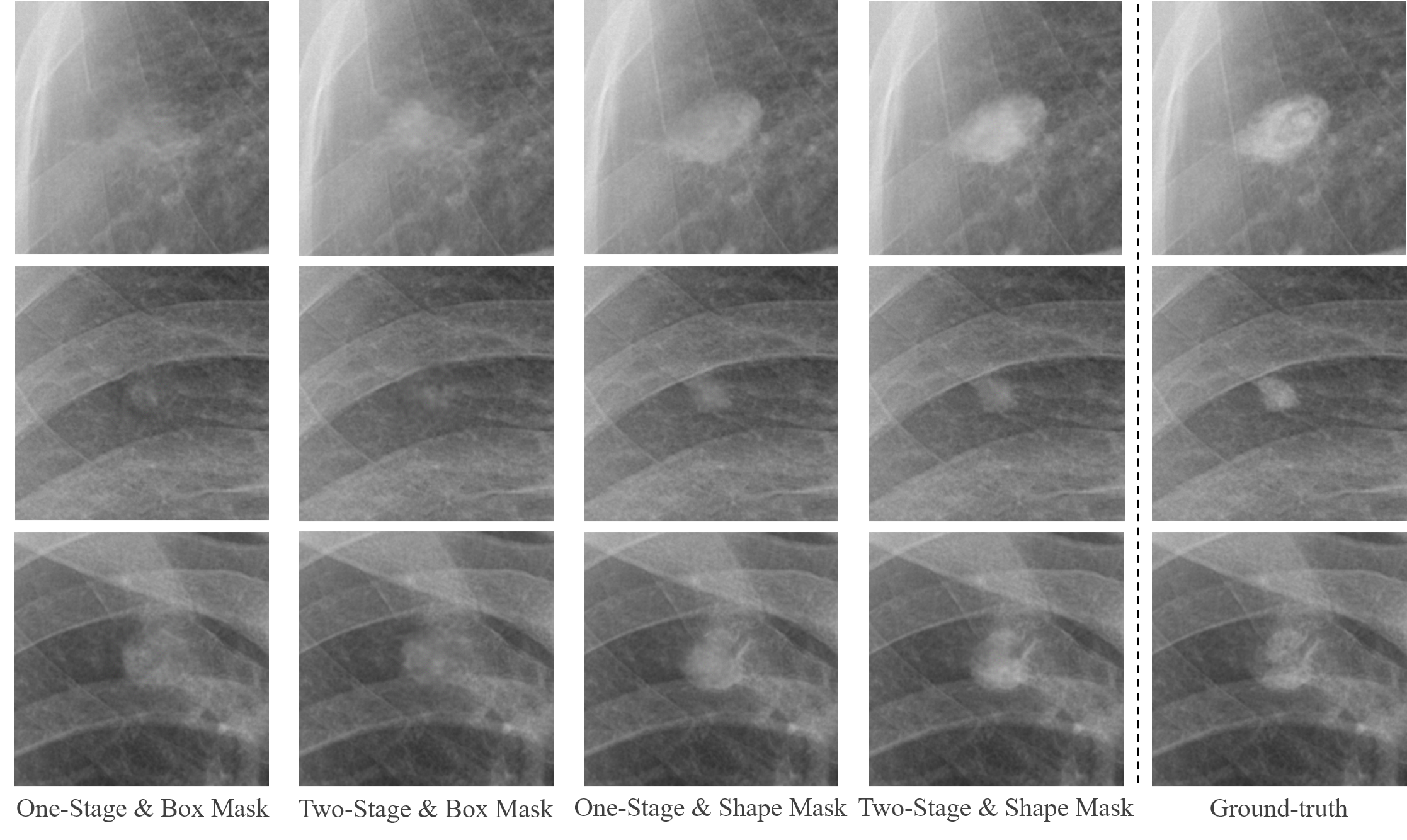}
    \caption{Visualization of the effects of using two-stage network architecture and conditional shape mask for lung nodule synthesis.}
    \label{fig:syn_ablation}
\end{figure*}

\begin{table*}[htbp]
\centering
\caption{Quantitative ablation study of different settings (↓: Lower is better; ↑: Higher is better).}
\label{tab:syn_ablation}
\setlength{\tabcolsep}{3mm}{
\begin{tabular}{cc|cccc|ccc}
\hline
 &  & \multicolumn{4}{c|}{Full Patch} & \multicolumn{3}{c}{Masked Region} \\ \cline{3-9} 
\multirow{-2}{*}{Shape Mask} & \multirow{-2}{*}{Two-Stage} & {MAE↓} & {PSNR↑} & {SSIM↑} & {FID↓} & {MAE↓} & {PSNR↑} & {SSIM↑} \\ \hline \hline
 &  & 0.0023 & 39.7072 & 0.9688 & 4.6675 & 0.0313 & 28.1213 & 0.7421 \\
 & $\surd$ & 0.0022 & 40.0569 & 0.9704 & 4.4333 & 0.0300 & 28.4711 & 0.7442 \\
$\surd$ &  & 0.0008 & 46.5973 & 0.9900 & 2.2798 & 0.0305 & 28.6721 & 0.7497 \\
$\surd$ & $\surd$ & \textbf{0.0007} & \textbf{47.0932} & \textbf{0.9903} & \textbf{2.1268} & \textbf{0.0296} & \textbf{28.9815} & \textbf{0.7555} \\ \hline
\end{tabular}}
\end{table*}

\subsubsection{Ablation Study for Texture Generator}
To verify the effects of applying conditional shape mask and two-stage (corresponding to coarse-to-fine above) network architecture in our Texture Generator $G_t$, we conduct four experimental settings: 
(1) a one-stage model conditioned on the box masks;
(2) a two-stage model conditioned on the box masks;
(3) a one-stage model conditioned on the shape masks;
(4) a two-stage model conditioned on the shape masks (the proposed Texture Generator).
All the models share the same loss functions and experimental setups.

\begin{itemize}
    \item \textbf{Effect of Shape Mask}. 
    As shown in Table \ref{tab:syn_ablation}, nodule synthesis based on the shape masks can significantly improve the performance in comparison with the box masks. 
    The visualization results in Fig. \ref{fig:syn_ablation} are also consistent with this quantitative comparison. 
    It is observed that the models conditioned on the shape masks (Column 3 \& 4) generate the nodules with much sharper boundaries and contrasts in comparison with the ones based on the box masks (Column 1 \& 2).
    The superiority of applying shape masks as conditions is clearly illustrated in comparison with using box masks.
    
    \item \textbf{Effect of Two-stage Architecture}. 
    Table \ref{tab:syn_ablation} demonstrates that the scores of the two-stage models are slightly improved compared with the one-stage models when using the same kind of conditional masks.
    As shown in Fig. \ref{fig:syn_ablation}, adopting a two-stage Texture Generator (Column 2 \& 4) adds more subtle details to the generated nodules, making the results more vivid and realistic than the outputs of the one-stage models (Column 1 \& 3). 
    This indicates that employing a two-stage architecture allows the model to better grasp fine-grained details in feature extraction, thus generating visually plausible nodule textures. 
    
\end{itemize}

\subsection{Data Augmentation Assessment}

In order to evaluate the detection performance augmented by the synthesized nodule CXR images, five-fold cross-validation experiments are conducted using Free-response Receiver Operating Characteristic (FROC) analysis as the quantitative metrics. 
To be specific, we calculate the Area under Curve (AUC) and NODE21 Score\footnote{\url{https://node21.grand-challenge.org/Details/}} according to the FROC curves. 
NODE21 Score is defined as the weighted sum of AUC and the sensitivity at 0.25 false positive rate, which is represented as $score=0.75\times AUC+0.25\times{Sen}_{0.25}$. 

We combine different types of synthesized data with 2,420 real nodule CXR images to finetune the pre-trained detectors and use the remaining 604 real images for a fair comparison between our methods and other approaches. 
And we regard the detectors trained with real data only as the baseline models. 
Considering that it is reasonable to use traditional data augmentation techniques in practice, we apply random shift and flip ($p=0.5$ for both) in all experiments.

For each detector, we firstly evaluate the data augmentation effectiveness of adding different quantities (including 0, 500, 1,000, 1,500, and 2,000) of synthesized CXR images using our method.
Table \ref{tab:da_num} shows that compared to the baselines, significant improvements can be made when adding the synthesized data into training. 
For Faster R-CNN, the AUC and NODE21 Score grow higher as the synthesized data quantity increases, and they reach the peak at 1,500 added images.
When the synthesized data increases to 2,000, the scores begin to decline but are still larger than the ones at 500 images. 
This is an expected phenomenon because adding too much synthesized data can dramatically change the inherent distribution of the original real data and brings biases into the training. 
As for RetinaNet, the AUC and NODE21 Score reach the top earlier than Faster R-CNN at 1,000 images because there is limited room for improvement when the baseline already fits the real data very well.

To validate the improvement of our HEM-based data augmentation strategy, we test four settings: the baseline, Syn\&Select \cite{shen2021nodule}, CT-to-CXR \cite{litjens2010simulation}, and our method. 
Specifically, we choose the best model for each detector when using our method (Faster R-CNN at 1,500 synthesized images and RetinaNet at 1,000 ones), and keep the synthesized data quantities in other settings consistent with ours. 
As shown in Table \ref{tab:da_compare}, all the detectors achieve the best performances using our augmentation strategy.
It is worth noting that CT-to-CXR requires real CT nodule patches to simulate nodule CXR images, but the detectors using our method outperforms the models augmented by CT-to-CXR.
This indicates that the nodules generated by our synthesis framework are more effective than simply projecting 3D CT nodules to 2D CXR nodules.
To sum up, the proposed HEM-based data augmentation strategy has a favorable effect on effectively boosting the performance of lung nodule detection, in conjunction with our lung nodule synthesis framework.

\begin{table*}[htbp]
\centering
\caption{Quantitative evaluation on the effectiveness of adding different quantities of synthesized data using the proposed data augmentation strategy.}
\label{tab:da_num}
\setlength{\tabcolsep}{6mm}{
\begin{tabular}{c|c|cc}
\hline
Detector & Data Quantity & AUC & NODE21 Score \\ \hline \hline
\multirow{5}{*}{Faster R-CNN} & 2,420 real & 0.9090 ± 0.0038 & 0.8673 ± 0.0028 \\
 & 2,420 real + 500 synthesized & 0.9318 ± 0.0180 & 0.8941 ± 0.0296 \\
 & 2,420 real + 1,000 synthesized & 0.9368 ± 0.0155 & 0.8967 ± 0.0299 \\
 & 2,420 real + 1,500 synthesized & \textbf{0.9385 ± 0.0143} & \textbf{0.8971 ± 0.0286} \\
 & 2,420 real + 2,000 synthesized & 0.9359 ± 0.0153 & 0.8935 ± 0.0290 \\ \hline
\multirow{5}{*}{RetinaNet} & 2,420 real & 0.9359 ± 0.0153 & 0.8935 ± 0.0290 \\
 & 2,420 real + 500 synthesized & 0.9653 ± 0.0027 & 0.9080 ± 0.0041 \\
 & 2,420 real + 1,000 synthesized & \textbf{0.9679 ± 0.0011} & \textbf{0.9097 ± 0.0035} \\
 & 2,420 real + 1,500 synthesized & 0.9617 ± 0.0024 & 0.9029 ± 0.0049 \\
 & 2,420 real + 2,000 synthesized & 0.9611 ± 0.0059 & 0.9020 ± 0.0037 \\ \hline
\end{tabular}}
\end{table*}

\begin{table*}[htbp]
\centering
\caption{Quantitative comparison between the proposed data augmentation strategy and other methods based on lung nodule synthesis.}
\label{tab:da_compare}
\setlength{\tabcolsep}{6mm}{
\begin{tabular}{c|c|cc}
\hline
Detector & Method & AUC & NODE21 Score \\ \hline \hline
\multirow{4}{*}{Faster R-CNN} & No Data Augmentation & 0.9090 ± 0.0038 & 0.8673 ± 0.0028 \\
 & Syn\&Select \cite{shen2021nodule} & 0.9129 ± 0.0057 & 0.8681 ± 0.0049 \\
 & CT-to-CXR \cite{litjens2010simulation} & 0.9176 ± 0.0019 & 0.8710 ± 0.0016 \\
 & \textbf{Ours} & \textbf{0.9385 ± 0.0143} & \textbf{0.8971 ± 0.0286} \\ \hline
\multirow{4}{*}{RetinaNet} & No Augmentation & 0.9359 ± 0.0153 & 0.8935 ± 0.0290 \\
 & Syn\&Select \cite{shen2021nodule} & 0.9584 ± 0.0054 & 0.9039 ± 0.0061 \\
 & CT-to-CXR \cite{litjens2010simulation} & 0.9626 ± 0.0015 & 0.9051 ± 0.0026 \\
 & \textbf{Ours} & \textbf{0.9679 ± 0.0011} & \textbf{0.9097 ± 0.0035} \\ \hline
\end{tabular}}
\end{table*}

\section{Conclusion and Limitation}
\label{sec:conclusion}
Existing deep learning-based CAD systems for lung nodule detection heavily rely on the large quantity and the diversity of well-annotated CXR images, which poses a great challenge on real-world medical data collection.
To overcome this problem, several methods have been developed to synthesize nodule CXR images for data augmentation.
In this paper, we propose a novel lung nodule synthesis framework based on image inpainting, which hierarchically decomposes the nodule synthesis procedure into Shape Generation, Size Modulation, and Texture Synthesis.
By applying quantitatively controllable shape masks as conditions, our synthesis framework provides diverse variations of nodule shapes, precise controls of nodule sizes, and visual plausibility of nodule textures.
Moreover, we further utilize the controllability of the proposed synthesis framework to design the HEM-based data augmentation strategy for effective improvement on nodule detection performance.

The evaluation on the image quality of the synthesized data demonstrates the key of the proposed lung nodule synthesis framework.
We firstly compare our method with SOTA image inpainting methods, and the experimental results show that our method achieves the most realistic visualization as well as the highest scores in all quantitative metrics.
Since obvious and distinctive structures hardly exist within a pulmonary nodule, the inpainting methods that incorporate structural information into modeling bring little helps but instability into image synthesis, leading to unsatisfactory results.
In contrast, our method benefits from the use of gated convolutions, which can dynamically select useful features from background contexts for visually plausible lung nodule synthesis.
In the ablation study, we further evaluate the effectiveness of the conditional shape mask and the two-stage architecture used in our method. 
The synthesized results turn out to be clearer and more realistic when conditioned on the shape masks, as against the blurry and diffuse outputs conditioned on the box masks.
Further, the comparison between one-stage and two-stage architectures illustrates that adopting a two-stage architecture benefits lung nodule synthesis as it makes the synthesized results look more vivid.
The repetitive utilization of the basic network in the two-stage architecture can extensively enlarges the receptive field of the model, thus refining local details as well as removing unfavorable blurs and artifacts.

As for data augmentation, we propose to synthesize training data according to the nodule attribute distribution which is established from missed nodules.
The core idea is that the detection performance can be effectively improved by feeding only a small number of hard training examples.
The comparison shows that our method outperforms other methods in augmenting nodule detection.
For one thing, our synthesis framework can achieve better image quality of the generated nodules.
For another, the pre-trained detector is more suitable for discriminating hard examples rather than an extra classifier trained from scratch in our previous work \cite{shen2021nodule}.
Moreover, the proposed data augmentation strategy is interpretable because the synthesized data are based on a particular nodule attribute which can be directly used as a diagnostic indicator.
Meanwhile, the comparison between our method and CT-to-CXR demonstrates that the visual plausibility of our synthesized nodules is comparable to that of the projected nodules using real CT templates because our method achieves higher scores in data augmentation.
This also suggests that our data augmentation strategy is more cost-effective as it can synthesize high-fidelity nodule CXR images without any extra CT data.

Our current study has several limitations that would be targeted in future works.
First, our lung nodule synthesis framework can automatically generate nodule textures based on the surrounding contexts but is unable to control the variation of its appearance and density.
The control of these attributes need to be further advanced in the future.
Second, it is difficult to quantitatively measure the image quality of the synthesized nodules since the scores are sometimes inconsistent with human perceptual judgment.
Third, the effectiveness of utilizing other attributes (\eg, shape or texture) in the proposed data augmentation strategy has not been investigated yet because quantitative and interpretable controls of these attributes are not supported in the current lung nodule synthesis framework.
It is worth conducting more studies for incorporating these attribute distributions into lung nodule synthesis as well as revealing the difference of data augmentation effectiveness when using different attributes.
Besides, our lung nodule synthesis framework is only applied in data augmentation so far. 
Its potential applications can be extended to other scenes such as counterfactual explanation for CAD systems and virtual reality for training radiologists in the future.

\bibliographystyle{IEEEtran}
\bibliography{ref.bib}

\begin{thebibliography}{10}
\providecommand{\url}[1]{#1}
\csname url@samestyle\endcsname
\providecommand{\newblock}{\relax}
\providecommand{\bibinfo}[2]{#2}
\providecommand{\BIBentrySTDinterwordspacing}{\spaceskip=0pt\relax}
\providecommand{\BIBentryALTinterwordstretchfactor}{4}
\providecommand{\BIBentryALTinterwordspacing}{\spaceskip=\fontdimen2\font plus
\BIBentryALTinterwordstretchfactor\fontdimen3\font minus
  \fontdimen4\font\relax}
\providecommand{\BIBforeignlanguage}[2]{{%
\expandafter\ifx\csname l@#1\endcsname\relax
\typeout{** WARNING: IEEEtran.bst: No hyphenation pattern has been}%
\typeout{** loaded for the language `#1'. Using the pattern for}%
\typeout{** the default language instead.}%
\else
\language=\csname l@#1\endcsname
\fi
#2}}
\providecommand{\BIBdecl}{\relax}
\BIBdecl

\bibitem{bray2018global}
F.~Bray, J.~Ferlay, I.~Soerjomataram, R.~L. Siegel, L.~A. Torre, and A.~Jemal,
  ``Global cancer statistics 2018: Globocan estimates of incidence and
  mortality worldwide for 36 cancers in 185 countries,'' \emph{CA: a cancer
  journal for clinicians}, vol.~68, no.~6, pp. 394--424, 2018.

\bibitem{siegel2019cancer}
R.~L. Siegel, K.~D. Miller, and A.~Jemal, ``Cancer statistics, 2019,''
  \emph{CA: a cancer journal for clinicians}, vol.~69, no.~1, pp. 7--34, 2019.

\bibitem{hansell2008fleischner}
D.~M. Hansell, A.~A. Bankier, H.~MacMahon, T.~C. McLoud, N.~L. Muller, and
  J.~Remy, ``Fleischner society: glossary of terms for thoracic imaging,''
  \emph{Radiology}, vol. 246, no.~3, pp. 697--722, 2008.

\bibitem{bhargavan2008trends}
M.~Bhargavan, ``Trends in the utilization of medical procedures that use
  ionizing radiation,'' \emph{Health physics}, vol.~95, no.~5, pp. 612--627,
  2008.

\bibitem{shah2003missed}
P.~K. Shah, J.~H. Austin, C.~S. White, P.~Patel, L.~B. Haramati, G.~D. Pearson,
  M.~C. Shiau, and Y.~M. Berkmen, ``Missed non--small cell lung cancer:
  radiographic findings of potentially resectable lesions evident only in
  retrospect,'' \emph{Radiology}, vol. 226, no.~1, pp. 235--241, 2003.

\bibitem{litjens2010simulation}
G.~J. Litjens, L.~Hogeweg, A.~M. Schilham, P.~A. de~Jong, M.~A. Viergever, and
  B.~v. Ginneken, ``Simulation of nodules and diffuse infiltrates in chest
  radiographs using ct templates,'' in \emph{International Conference on
  Medical Image Computing and Computer-Assisted Intervention}.\hskip 1em plus
  0.5em minus 0.4em\relax Springer, 2010, pp. 396--403.

\bibitem{nam2019development}
J.~G. Nam, S.~Park, E.~J. Hwang, J.~H. Lee, K.-N. Jin, K.~Y. Lim, T.~H. Vu,
  J.~H. Sohn, S.~Hwang, J.~M. Goo \emph{et~al.}, ``Development and validation
  of deep learning--based automatic detection algorithm for malignant pulmonary
  nodules on chest radiographs,'' \emph{Radiology}, vol. 290, no.~1, pp.
  218--228, 2019.

\bibitem{li2020multi}
X.~Li, L.~Shen, X.~Xie, S.~Huang, Z.~Xie, X.~Hong, and J.~Yu,
  ``Multi-resolution convolutional networks for chest x-ray radiograph based
  lung nodule detection,'' \emph{Artificial intelligence in medicine}, vol.
  103, p. 101744, 2020.

\bibitem{sim2020deep}
Y.~Sim, M.~J. Chung, E.~Kotter, S.~Yune, M.~Kim, S.~Do, K.~Han, H.~Kim,
  S.~Yang, D.-J. Lee \emph{et~al.}, ``Deep convolutional neural network--based
  software improves radiologist detection of malignant lung nodules on chest
  radiographs,'' \emph{Radiology}, vol. 294, no.~1, pp. 199--209, 2020.

\bibitem{wang2021realistic}
Q.~Wang, X.~Zhang, W.~Zhang, M.~Gao, S.~Huang, J.~Wang, J.~Zhang, D.~Yang, and
  C.~Liu, ``Realistic lung nodule synthesis with multi-target co-guided
  adversarial mechanism,'' \emph{IEEE Transactions on Medical Imaging},
  vol.~40, no.~9, pp. 2343--2353, 2021.

\bibitem{cunniff2000informed}
C.~Cunniff, J.~L. Byrne, L.~M. Hudgins, J.~B. Moeschler, A.~H. Olney, R.~M.
  Pauli, L.~H. Seaver, C.~A. Stevens, and C.~Figone, ``Informed consent for
  medical photographs,'' \emph{Genetics in Medicine}, vol.~2, no.~6, pp.
  353--355, 2000.

\bibitem{yi2019generative}
X.~Yi, E.~Walia, and P.~Babyn, ``Generative adversarial network in medical
  imaging: A review,'' \emph{Medical image analysis}, vol.~58, p. 101552, 2019.

\bibitem{li2021medical}
X.~Li, Y.~Jiang, J.~J. Rodriguez-Andina, H.~Luo, S.~Yin, and O.~Kaynak, ``When
  medical images meet generative adversarial network: recent development and
  research opportunities,'' \emph{Discover Artificial Intelligence}, vol.~1,
  no.~1, pp. 1--20, 2021.

\bibitem{shorten2019survey}
C.~Shorten and T.~M. Khoshgoftaar, ``A survey on image data augmentation for
  deep learning,'' \emph{Journal of big data}, vol.~6, no.~1, pp. 1--48, 2019.

\bibitem{ann2021multi}
K.~Ann, Y.~Jang, H.~Shim, and H.-J. Chang, ``Multi-scale conditional generative
  adversarial network for small-sized lung nodules using class activation
  region influence maximization,'' \emph{IEEE Access}, vol.~9, pp.
  139\,426--139\,437, 2021.

\bibitem{gundel2020extracting}
S.~G{\"u}ndel, A.~A. Setio, S.~Grbic, A.~Maier, and D.~Comaniciu, ``Extracting
  and leveraging nodule features with lung inpainting for local feature
  augmentation,'' in \emph{International Workshop on Machine Learning in
  Medical Imaging}.\hskip 1em plus 0.5em minus 0.4em\relax Springer, 2020, pp.
  504--512.

\bibitem{shen2021nodule}
Z.~Shen, X.~Ouyang, Z.~Wang, Y.~Zhan, Z.~Xue, Q.~Wang, J.-Z. Cheng, and
  D.~Shen, ``Nodule synthesis and selection for augmenting chest x-ray nodule
  detection,'' in \emph{Chinese Conference on Pattern Recognition and Computer
  Vision (PRCV)}.\hskip 1em plus 0.5em minus 0.4em\relax Springer, 2021, pp.
  536--547.

\bibitem{goodfellow2014generative}
I.~Goodfellow, J.~Pouget-Abadie, M.~Mirza, B.~Xu, D.~Warde-Farley, S.~Ozair,
  A.~Courville, and Y.~Bengio, ``Generative adversarial nets,'' \emph{Advances
  in neural information processing systems}, vol.~27, 2014.

\bibitem{elharrouss2020image}
O.~Elharrouss, N.~Almaadeed, S.~Al-Maadeed, and Y.~Akbari, ``Image inpainting:
  A review,'' \emph{Neural Processing Letters}, vol.~51, no.~2, pp. 2007--2028,
  2020.

\bibitem{shrivastava2016training}
A.~Shrivastava, A.~Gupta, and R.~Girshick, ``Training region-based object
  detectors with online hard example mining,'' in \emph{Proceedings of the IEEE
  conference on computer vision and pattern recognition}, 2016, pp. 761--769.

\bibitem{bertalmio2000image}
M.~Bertalmio, G.~Sapiro, V.~Caselles, and C.~Ballester, ``Image inpainting,''
  in \emph{Proceedings of the 27th annual conference on Computer graphics and
  interactive techniques}, 2000, pp. 417--424.

\bibitem{efros2001image}
A.~A. Efros and W.~T. Freeman, ``Image quilting for texture synthesis and
  transfer,'' in \emph{Proceedings of the 28th annual conference on Computer
  graphics and interactive techniques}, 2001, pp. 341--346.

\bibitem{kwatra2005texture}
V.~Kwatra, I.~Essa, A.~Bobick, and N.~Kwatra, ``Texture optimization for
  example-based synthesis,'' in \emph{ACM SIGGRAPH 2005 Papers}, 2005, pp.
  795--802.

\bibitem{barnes2009patchmatch}
C.~Barnes, E.~Shechtman, A.~Finkelstein, and D.~B. Goldman, ``Patchmatch: A
  randomized correspondence algorithm for structural image editing,'' \emph{ACM
  Trans. Graph.}, vol.~28, no.~3, p.~24, 2009.

\bibitem{liu2018image}
G.~Liu, F.~A. Reda, K.~J. Shih, T.-C. Wang, A.~Tao, and B.~Catanzaro, ``Image
  inpainting for irregular holes using partial convolutions,'' in
  \emph{Proceedings of the European conference on computer vision (ECCV)},
  2018, pp. 85--100.

\bibitem{yu2019free}
J.~Yu, Z.~Lin, J.~Yang, X.~Shen, X.~Lu, and T.~S. Huang, ``Free-form image
  inpainting with gated convolution,'' in \emph{Proceedings of the IEEE/CVF
  International Conference on Computer Vision}, 2019, pp. 4471--4480.

\bibitem{nazeri2019edgeconnect}
K.~Nazeri, E.~Ng, T.~Joseph, F.~Qureshi, and M.~Ebrahimi, ``Edgeconnect:
  Structure guided image inpainting using edge prediction,'' in
  \emph{Proceedings of the IEEE/CVF International Conference on Computer Vision
  Workshops}, 2019, pp. 0--0.

\bibitem{ren2019structureflow}
Y.~Ren, X.~Yu, R.~Zhang, T.~H. Li, S.~Liu, and G.~Li, ``Structureflow: Image
  inpainting via structure-aware appearance flow,'' in \emph{Proceedings of the
  IEEE/CVF International Conference on Computer Vision}, 2019, pp. 181--190.

\bibitem{li2019progressive}
J.~Li, F.~He, L.~Zhang, B.~Du, and D.~Tao, ``Progressive reconstruction of
  visual structure for image inpainting,'' in \emph{Proceedings of the IEEE/CVF
  International Conference on Computer Vision}, 2019, pp. 5962--5971.

\bibitem{liu2020rethinking}
H.~Liu, B.~Jiang, Y.~Song, W.~Huang, and C.~Yang, ``Rethinking image inpainting
  via a mutual encoder-decoder with feature equalizations,'' in \emph{European
  Conference on Computer Vision}.\hskip 1em plus 0.5em minus 0.4em\relax
  Springer, 2020, pp. 725--741.

\bibitem{guo2021image}
X.~Guo, H.~Yang, and D.~Huang, ``Image inpainting via conditional texture and
  structure dual generation,'' in \emph{Proceedings of the IEEE/CVF
  International Conference on Computer Vision}, 2021, pp. 14\,134--14\,143.

\bibitem{ost2012decision}
D.~E. Ost and M.~K. Gould, ``Decision making in patients with pulmonary
  nodules,'' \emph{American journal of respiratory and critical care medicine},
  vol. 185, no.~4, pp. 363--372, 2012.

\bibitem{salehinejad2018synthesizing}
H.~Salehinejad, E.~Colak, T.~Dowdell, J.~Barfett, and S.~Valaee, ``Synthesizing
  chest x-ray pathology for training deep convolutional neural networks,''
  \emph{IEEE transactions on medical imaging}, vol.~38, no.~5, pp. 1197--1206,
  2018.

\bibitem{radford2015unsupervised}
A.~Radford, L.~Metz, and S.~Chintala, ``Unsupervised representation learning
  with deep convolutional generative adversarial networks,'' \emph{arXiv
  preprint arXiv:1511.06434}, 2015.

\bibitem{karras2017progressive}
T.~Karras, T.~Aila, S.~Laine, and J.~Lehtinen, ``Progressive growing of gans
  for improved quality, stability, and variation,'' \emph{arXiv preprint
  arXiv:1710.10196}, 2017.

\bibitem{sogancioglu2018chest}
E.~Sogancioglu, S.~Hu, D.~Belli, and B.~van Ginneken, ``Chest x-ray inpainting
  with deep generative models,'' \emph{arXiv preprint arXiv:1809.01471}, 2018.

\bibitem{pathak2016context}
D.~Pathak, P.~Krahenbuhl, J.~Donahue, T.~Darrell, and A.~A. Efros, ``Context
  encoders: Feature learning by inpainting,'' in \emph{Proceedings of the IEEE
  conference on computer vision and pattern recognition}, 2016, pp. 2536--2544.

\bibitem{yeh2017semantic}
R.~A. Yeh, C.~Chen, T.~Yian~Lim, A.~G. Schwing, M.~Hasegawa-Johnson, and M.~N.
  Do, ``Semantic image inpainting with deep generative models,'' in
  \emph{Proceedings of the IEEE conference on computer vision and pattern
  recognition}, 2017, pp. 5485--5493.

\bibitem{yu2018generative}
J.~Yu, Z.~Lin, J.~Yang, X.~Shen, X.~Lu, and T.~S. Huang, ``Generative image
  inpainting with contextual attention,'' in \emph{Proceedings of the IEEE
  conference on computer vision and pattern recognition}, 2018, pp. 5505--5514.

\bibitem{ronneberger2015u}
O.~Ronneberger, P.~Fischer, and T.~Brox, ``U-net: Convolutional networks for
  biomedical image segmentation,'' in \emph{International Conference on Medical
  image computing and computer-assisted intervention}.\hskip 1em plus 0.5em
  minus 0.4em\relax Springer, 2015, pp. 234--241.

\bibitem{jegou2017one}
S.~J{\'e}gou, M.~Drozdzal, D.~Vazquez, A.~Romero, and Y.~Bengio, ``The one
  hundred layers tiramisu: Fully convolutional densenets for semantic
  segmentation,'' in \emph{Proceedings of the IEEE conference on computer
  vision and pattern recognition workshops}, 2017, pp. 11--19.

\bibitem{mao2017least}
X.~Mao, Q.~Li, H.~Xie, R.~Y. Lau, Z.~Wang, and S.~Paul~Smolley, ``Least squares
  generative adversarial networks,'' in \emph{Proceedings of the IEEE
  international conference on computer vision}, 2017, pp. 2794--2802.

\bibitem{goetschalckx2019ganalyze}
L.~Goetschalckx, A.~Andonian, A.~Oliva, and P.~Isola, ``Ganalyze: Toward visual
  definitions of cognitive image properties,'' in \emph{Proceedings of the
  ieee/cvf international conference on computer vision}, 2019, pp. 5744--5753.

\bibitem{shen2020interpreting}
Y.~Shen, J.~Gu, X.~Tang, and B.~Zhou, ``Interpreting the latent space of gans
  for semantic face editing,'' in \emph{Proceedings of the IEEE/CVF Conference
  on Computer Vision and Pattern Recognition}, 2020, pp. 9243--9252.

\bibitem{isola2017image}
P.~Isola, J.-Y. Zhu, T.~Zhou, and A.~A. Efros, ``Image-to-image translation
  with conditional adversarial networks,'' in \emph{Proceedings of the IEEE
  conference on computer vision and pattern recognition}, 2017, pp. 1125--1134.

\bibitem{miyato2018spectral}
T.~Miyato, T.~Kataoka, M.~Koyama, and Y.~Yoshida, ``Spectral normalization for
  generative adversarial networks,'' \emph{arXiv preprint arXiv:1802.05957},
  2018.

\bibitem{johnson2016perceptual}
J.~Johnson, A.~Alahi, and L.~Fei-Fei, ``Perceptual losses for real-time style
  transfer and super-resolution,'' in \emph{European conference on computer
  vision}.\hskip 1em plus 0.5em minus 0.4em\relax Springer, 2016, pp. 694--711.

\bibitem{simonyan2014very}
K.~Simonyan and A.~Zisserman, ``Very deep convolutional networks for
  large-scale image recognition,'' \emph{arXiv preprint arXiv:1409.1556}, 2014.

\bibitem{deng2009imagenet}
J.~Deng, W.~Dong, R.~Socher, L.-J. Li, K.~Li, and L.~Fei-Fei, ``Imagenet: A
  large-scale hierarchical image database,'' in \emph{2009 IEEE conference on
  computer vision and pattern recognition}.\hskip 1em plus 0.5em minus
  0.4em\relax Ieee, 2009, pp. 248--255.

\bibitem{paszke2019pytorch}
A.~Paszke, S.~Gross, F.~Massa, A.~Lerer, J.~Bradbury, G.~Chanan, T.~Killeen,
  Z.~Lin, N.~Gimelshein, L.~Antiga \emph{et~al.}, ``Pytorch: An imperative
  style, high-performance deep learning library,'' \emph{Advances in neural
  information processing systems}, vol.~32, 2019.

\bibitem{kingma2014adam}
D.~P. Kingma and J.~Ba, ``Adam: A method for stochastic optimization,''
  \emph{arXiv preprint arXiv:1412.6980}, 2014.

\bibitem{ren2015faster}
S.~Ren, K.~He, R.~Girshick, and J.~Sun, ``Faster r-cnn: Towards real-time
  object detection with region proposal networks,'' \emph{Advances in neural
  information processing systems}, vol.~28, 2015.

\bibitem{lin2017focal}
T.-Y. Lin, P.~Goyal, R.~Girshick, K.~He, and P.~Doll{\'a}r, ``Focal loss for
  dense object detection,'' in \emph{Proceedings of the IEEE international
  conference on computer vision}, 2017, pp. 2980--2988.

\bibitem{he2016deep}
K.~He, X.~Zhang, S.~Ren, and J.~Sun, ``Deep residual learning for image
  recognition,'' in \emph{Proceedings of the IEEE conference on computer vision
  and pattern recognition}, 2016, pp. 770--778.

\bibitem{heusel2017gans}
M.~Heusel, H.~Ramsauer, T.~Unterthiner, B.~Nessler, and S.~Hochreiter, ``Gans
  trained by a two time-scale update rule converge to a local nash
  equilibrium,'' \emph{Advances in neural information processing systems},
  vol.~30, 2017.

\end{thebibliography}

\end{document}